\documentclass[9pt,twocolumn,twoside]{pnas-new}

\templatetype{pnasresearcharticle} 
\usepackage{amsmath,amssymb}
\usepackage{mathtools}
\usepackage{arydshln}
\usepackage{physics}
\usepackage{bm}

\usepackage{color}

\providecommand{\diff}[1]{{#1}} 

\title{Formation of human kinship structures depending on population size and cultural mutation rate}
\author[a, b, c]{Kenji Itao}
\author[d, e, 1]{Kunihiko Kaneko} 

\affil[a]{Department of Basic Science, Graduate School of Arts and Sciences, University of Tokyo, Komaba 3-8-1, Meguro, Tokyo 153-8902, Japan.}
\affil[b]{Computational Group Dynamics Collaboration Unit, RIKEN Center for Brain Science, 2-1 Hirosawa, Wako, Saitama 351-0198, Japan.}
\affil[c]{\textit{BirthRites} Lise Meitner Research Group, Max Planck Institute for Evolutionary Anthropology, Deutscher Platz 6, Leipzig, 04103, Germany}
\affil[d]{The Niels Bohr Institute, University of Copenhagen, Blegdamsvej 17, Copenhagen, 2100-DK, Denmark.}
\affil[e]{Universal Biology Institute, University of Tokyo, Komaba 3-8-1, Meguro, Tokyo 153-8902, Japan.}

\leadauthor{Itao} 

\significancestatement{
\diff{Kinship structures with rules of marriage among cultural groups are the basis of social relationships in many traditional human societies. Cultural anthropologists have identified typical classes in worldwide structures with cyclic exchange of mates among clans. By modeling anthropologically observed kinship interactions including sibling unity, in-law ties, and sexual rivalry and simulating multilevel evolution, we demonstrate the spontaneous emergence of these structures marked by self-organized interdependence among differentiated groups. Our numerical results and analytical estimates show how kinship structures, in particular the cycle lengths of marriage exchange and the number of distinct cycles within society, depend on the population size and cultural mutation rate. This study offers mechanisms for the evolution of complex social structures from a statistical physics perspective.}
}

\authorcontributions{K.I. and K.K. designed the model; K.I. conducted the simulations; K.I. and K.K. analyzed the data; and K.I. and K.K. wrote the paper.}
\authordeclaration{The authors declare no conflict of interest.}
\correspondingauthor{\textsuperscript{1} To whom correspondence should be addressed. Email: kaneko@complex.c.u-tokyo.ac.jp}


\begin{abstract}
How does social complexity depend on population size and cultural transmission? Kinship structures in traditional societies provide a fundamental illustration, where cultural rules between clans determine people's marriage possibilities. Here we propose a simple model of kinship interactions that considers kin and in-law cooperation and sexual rivalry. In this model, multiple societies compete. Societies consist of multiple families with different cultural traits and mating preferences. These values determine interactions and hence the growth rate of families and are transmitted to offspring with mutations. Through a multilevel evolutionary simulation, family traits and preferences are grouped into multiple clans with inter-clan mating preferences. It illustrates the emergence of kinship structures as the spontaneous formation of interdependent cultural associations. Emergent kinship structures are characterized by the cycle length of marriage exchange and the number of cycles in society. We numerically and analytically clarify their parameter dependence. The relative importance of cooperation versus rivalry determines whether attraction or repulsion exists between families. Different structures evolve as locally stable attractors. The probabilities of formation and collapse of complex structures depend on the number of families and the mutation rate, showing characteristic scaling relationships. It is now possible to explore macroscopic kinship structures based on microscopic interactions, together with their environmental dependence and the historical causality of their evolution. We propose the basic causal mechanism of the formation of typical human social structures by referring to ethnographic observations and concepts from statistical physics and multilevel evolution. Such interdisciplinary collaboration will unveil universal features in human societies.
\end{abstract}

\dates{This manuscript was compiled on \today}
\doi{\url{www.pnas.org/cgi/doi/10.1073/pnas.XXXXXXXXXX}}

\begin{document}
\maketitle
\thispagestyle{firststyle}
\ifthenelse{\boolean{shortarticle}}{\ifthenelse{\boolean{singlecolumn}}{\abscontentformatted}{\abscontent}}{}

\firstpage[4]{3}

\dropcap{S}ocial complexity, reflecting the scale of society and the degree of functional differentiation and hierarchy, has increased in human history \cite{service1962primitive, feinman2013emergence, turchin2018quantitative}.
To discuss the conditions for the increase in complexity, cultural evolutionists have discussed the necessary population size $N$ and mutation rate $\mu$ for a stable increase in technological levels \cite{henrich2004demography, powell2009late}. However, there is another type of complexity characterized by the interdependency of differentiated groups.
In many traditional human societies, people are classified into several cultural peer groups called clans, and customs determine the possibility of marriage depending on the clan to which they belong \cite{levi1969elementary, fox1983kinship}. Here, rules often promote exogamy, i.e., the marriage with different clan members. Relationships between clans are interdependent in that differentiated groups need each other for reproduction.
Such interdependent group structures regarding marriage are termed kinship structures and are considered one of the oldest and most frequent human social organizations \cite{service1962primitive, fox1983kinship, shenk2011rebirth}.

Such interdependent structures, composed of interacting differentiated elements where the survival of the parts necessitates the existence of the whole, have been observed in many complex biological systems. Examples include the interdependency between replicating molecules \cite{takeuchi2017origin}, that between gametes \cite{zadorin2023multilevel}, and that between social insects \cite{nowak2010evolution}. Their origins are discussed by focusing on the interactions between elements, and the adaptivity of the entire system. Thus, multilevel evolution is a useful framework that considers both element- and group-level adaptivity simultaneously \cite{hogeweg1994multilevel}. 
In this study, we investigate the origins of kinship structures by modeling people's interactions based on ethnographic observations and simulating the multilevel evolution of families and societies. Therefore, we explore a generalizable mechanism for interdependent group formation, together with its dependence on $N$ and $\mu$.

Kinship interactions include the unity of siblings, in-law bonds, and sexual rivalry \cite{levi1969elementary, leach1982social, alvard2003kinship, alvard2011genetic, thomas2018kinship}.
Anthropologists have proposed ``descent theory'' and ``alliance theory'' to explain kinship structures. Clan members united by shared descent cooperate in daily activities \cite{levi1969elementary, alvard2003kinship, alvard2011genetic, thomas2018kinship}. Also, the cooperation between in-laws is repeatedly observed in diverse regions \citep{leach1982social, shenk2011rebirth, power2019cooperation}. 
Here, ``cooperation\footnote{\diff{We adopted the definition of ``cooperation'' as a behavior that benefits the recipient but not necessarily the actor (i.e., including mutual benefit and altruism) \cite{sachs2004evolution, west2007social}. We are aware that, especially in the game theoretic context, cooperation is often defined as a behavior where one individual pays a cost for another to receive a benefit \cite{rand2013human}, and the evolution of cooperation is an important issue. Here, we are not interested in how cooperative relationships evolve or are maintained. Rather, we focus on how they are relevant to forming social structures.}}'' refers to any form of help for survival, including co-hunting, co-fighting, labor, and payment \cite{leach1954political, alvard2003kinship, alvard2009kinship, alvard2011genetic}. The cooperation between in-laws can be bidirectional, a brideprice from the groom's family to the bride's, or a dowry in the opposite direction \citep{leach1982social, anderson2007economics, shenk2011rebirth}. Such cooperation is observed in many societies including highland Burmese society where the bride’s family, after sending her off to marry, can request gifts and labor from the groom's family \cite{leach1954political}.
In the Standard Cross-Cultural Sample (SCCS) database, 60\% of societies engage in brideprice transactions at marriage, while 5\% practice dowry, and 13\% exhibit bidirectional exchanges (see the variable 1195) \cite{murdock1969standard, kirby2016d}. When considering other forms of cooperation as well, more societies would perform in-law cooperation \cite{shenk2011rebirth, alvard2003kinship, alvard2009kinship}.

Between rival families competing for mates, the field studies also revealed strong conflicts that can be fatal \citep{Helbling1999, chagnon1988life, blainey1976triumph, otterbein1997origins}. Sexual jealousy is considered one of the main causes of homicide worldwide \cite{daly1982male, chagnon1988life, ferguson2000causes, buss2013sexual}. 
In particular, in Yanomamo society, kin and in-laws cooperate in raids that are motivated by sexual jealousy \cite{chagnon1988life, alvard2009kinship}. Indeed, intra-sexual aggression has been reported to be stronger when the sex ratio in the experiment is biased, but cross-cultural comparisons have shown a small correlation between sexual rivalry and sex ratio bias, and sexual rivalry exists even when the sex ratio is equal \cite{schacht2014too, moss2016biased}. Additionally, the SCCS data (variables 964, 965, 983, 1136) and the ethnographic reports suggest that married individuals who commit adultery, as well as adulterers, can be punished by death or divorce and are obliged to pay compensation cross-culturally \cite{daly1982male, murdock1969standard, musharbash2010marriage, kirby2016d}.
In this context, marriage rules that distinguish between preferable and unpreferable individuals are viewed as ``laws'' to avert conflicts resulting from reproductive pursuits \cite{chagnon2017manipulating}.

Although kin unity and sexual rivalry are observed in many non-human animals, no non-human apes recognize the affinal network or have bonds with in-laws \cite{chapais2009primeval}. This difference in cognitive ability purportedly allows for the complex kinship systems in human societies \cite{planer2021towards}; consequently, the study of the origins of kinship systems is fundamental to understanding human history.

Anthropologists have observed that in traditional small-scale societies, where these interactions matter, various kinship structures are organized \cite{fox1983kinship, shenk2011rebirth}.
In some cases, the rules merely prohibit marriage within clans because clan members are cultural siblings, even though they are not necessarily genetically related. In other cases, the rules further specify which clan members one is allowed to marry \cite{levi1969elementary}. In this study, we focus on such societies in which norms specify preferable mate candidates. In many societies, there is only one class of relatives who are legitimate potential mates \cite{chagnon2017manipulating, romney1958simplified, kouno2008algebraic}. Especially, cross-cousin (the mother's brother's daughter or father's sister's daughter) marriage is frequently promoted \cite{levi1969elementary, shenk2016consanguineous, chapais2014complex}.
By focusing on people's association with clans, the rules can be expressed simply: Let us consider the rule of matrilineal cross-cousin marriage and assume that children belong to the same clan as their fathers. If a man from clan $X$ marries a woman from clan $Y$ and they have a boy (Ego) in $X$, his mother originally belonged to $Y$ and her brother's children are in $Y$. Thus, the marriage preference for the mother's brother's daughters is expressed as a rule that men from clan $X$ should marry women in clan $Y$ in every generation. The set of such rules between clans represents the kinship structure.

Several tiers of kinship structures have been identified \cite{levi1969elementary, leach1954political}. In some societies, there are only two clans: $X$ and $Y$. People must choose mates from opposing clans. 
This structure is represented as $X\Leftrightarrow Y$, denoting the rule for women in clan $X$ to marry men in clan $Y$ as $X \Rightarrow Y$. This is called dual organization. Similarly, generalized exchange is characterized by a cyclic preference and represented as $X \Rightarrow Y \Rightarrow Z \Rightarrow \cdots \Rightarrow X$.
This structure collocates with the rule that the wife's clan should be different from the sister's husband's clan.
In cultural anthropology, structures are characterized by the number of clans constituting the cycle \cite{levi1969elementary, white1963anatomy}. The cycle length of the marriage relationships of clans is termed as marriage cycle length $c_m$.
$c_m = 1$ characterizes endogamy and $c_m \ge 2$ indicates exogamy.
Exogamy requires the interdependence of clans, in that each clan requires another to find mates.

Mathematical studies have analyzed kinship structures using group theory \cite{weil1949algebra, white1963anatomy, bush1963algebraic}. Ethnographic data analyses have revealed patterns of kinship terminology, that is, the categorization of relatives by ego, which presumably reflects kinship structures, and the transition between them \cite{allen1989evolution, passmore2021kin, passmore2023kinbank}. However, how such structural patterns are organized and what explains their diversity remain unclear.
To explore the evolution of kinship structures, the relationship between microscopic kinship interactions and macroscopic interdependent structures needs to be clarified. To this end, a statistical physics approach based on a simple model to demonstrate the emergence of macroscopic structures is useful and can explain the universality of kinship structures worldwide.

In this study, we hypothesize that the unity of siblings and bonds with in-laws drive assimilation and that sexual rivalry accelerates the dissimilation of families.
Previously, we numerically illustrated the evolution of various kinship structures depending on the strength of marital ties and rivalry, both theoretically and empirically \cite{itao2020evolution, itao2022emergence}. However, in multilevel evolution and the study of social complexity, the dependence on the number of units and mutation rate is essential \cite{takeuchi2017origin, traulsen2006evolution}.
In contrast, the previous numerical model considered trait inheritance from both parents, making it complicated to quantitatively analyze such parameter dependence. Here, we introduce a simplified model by considering unilineal inheritance only to reveal the essential mechanisms underlying the evolution of kinship structures and the scaling law in parameter dependence. Indeed, societies with unilineal descent systems are the majority in the ethnographic database \cite{murdock1969standard, kirby2016d}. Thus, the current simplified model will have anthropological significance.

With this simplified model, we demonstrate the evolution of various structures depending on the ratio of the strength of sexual rivalry to marital ties as well as the number of units and the mutation rate. We then, numerically and theoretically, scrutinize the conditions for the evolution of each structure. 

\section*{Model}
In the model, we consider a set of $S$ societies, each comprising several families that interact with each other. The number of families in a society is initially $N$ and changes through interactions. When the population reaches $2N$, the society splits into two, and families are randomly allocated to two daughter societies. Simultaneously, another society is randomly removed. Thus, the number of families in each society fluctuates between $0$ and $2N$, whereas the number of societies in the entire system remains fixed at $S$. In this way, multilevel evolution is introduced, where families and societies that grow faster replace others. This multilevel evolutionary framework has been widely adopted in biological and social evolutionary studies to explore the origins of group-level structures \cite{spencer2001multilevel, traulsen2006evolution, turchin2009evolution, takeuchi2017origin}.

Each family consists of a couple and children. It contains the parameters of the cultural trait $t$ and mating preference $p$. Traits represent any social feature by which people measure their cultural similarities, even without genetic relatedness, such as names, occupations, or totems \cite{levi1962pensee, sperber2004cognitive}. Preference denotes the preferable trait of a groom. These values determine interactions between families, including marriage, cooperation, and rivalry. The possibility of marriage between men in family $i$ and women in family $j$ is large when $t_i$ and $p_j$ are similar. We define this possibility as $\exp(-(t_i - p_j)^2)$. Then, the possibility of a man in the family $i$ finding a mate is given by the degree of preferredness: 
\begin{equation}
    D_p^i = \ev{\exp(-(t_i - p_j) ^2)}_j / \ev{\exp(-(t_i - t_j) ^2)}_j,
\end{equation}
which is the ratio of the number of women who prefer someone like him to the number of men like him. Here, $\ev{\cdot}_j$ represents the expected value averaged over families in the same society with varying $j$.

The number of children who survive to reproductive age is determined by the density of cooperators and rivals in the same society. Here, we consider cooperation between siblings and in-laws. Thus, family $j$ will be more likely to cooperate with family $i$ if $|t_i - t_j|$ or $|p_i - t_j|$ is small (i.e., if $i$ is a cultural sibling or potential mate for $j$\footnote{Note that, when $|p_i - t_j|$ is small, $i$ gets cooperation from $j$ because $i$ helps $j$'s reproduction, representing the brideprice-type in-law cooperation. The simulation results of model variants with the dowry-type and bidirectional cooperation are presented in Figs. S1 and S2, respectively.}). We define the degree of cooperation that $i$ receives from $j$ as $\exp(-(t_i - t_j) ^2) + (1-\exp(-(t_i - t_j) ^2))\exp(-(p_i - t_j) ^2)$ $(\simeq \exp(-\min(|t_i - t_j|, |p_i - t_j|) ^2))$.
Then, the density of cooperative friends for family $i$ is given by
\begin{align}
     D_f^i =& \langle \exp(-(t_i - t_j) ^2)\\ \nonumber
     &+ (1-\exp(-(t_i - t_j) ^2))\exp(-(p_i - t_j) ^2)\rangle_j. 
\end{align}
Although having affinity in addition to consanguinity could further increase the level of cooperation, according to the field study, there is a negative interaction between the two, and the level of cooperation does not increase as much as the sum of the two \cite{thomas2018kinship}. Therefore, we defined the above measure as the maximum value of cooperation by consanguinity and affinity.

Similarly, the density of competing rivals for family $i$ is given by\footnote{Depending on the society, various forms of sexual rivalry are observed. However, they are mostly caused by intra-sexual preference overlaps, and rules to avoid rivalry have been developed. The current formulation that overlapping female preferences lead to lower fitness can be interpreted as indicating that adulterous women are punished by death or are obliged to pay compensation. This lowers the fitness of these women's families. To express the ethnographic observation of male sexual violence, we also built a model variant in which men prefer women and overlapping male preferences lead to sexual rivalry, whose results are shown in Fig. S3.}
\begin{equation}
    D_r^i = \ev{\exp(-(p_i - p_j) ^2)}_j.
\end{equation}
Recall that the rules determining marriageable mates are viewed as ``laws'' to avert conflicts arising from reproductive striving \cite{chagnon2017manipulating}. This $D_r^i$ measures the possibility of sexual rivalry caused by jealousy, whereas $D_p^i$ assesses the marriage possibility based on the sex ratio of marriageable mates. To illustrate the significance of these measures, consider two extreme cases. In one scenario, all families share identical $t$ and $p$ values, resulting in both sexual rivalry and marriage possibilities being $1$. In the other scenario, each family has unique $t$ and $p$ values perfectly matched one-to-one; here, the possibility of marriage remains $1$, but that of sexual rivalry drops to $0$ due to strict laws to avoid rivalry.

Mortality increases when there are few cooperators and many rivals.
The children’s survival rate is then given by $D_f^i + \rho (1 - D_r^i)$, where $\rho$ denotes the relative importance of avoiding rivalry versus having cooperation.
Overall, the population growth rate of family $i$ is proportional to 
\begin{equation}
    g_i = (D_f^i + \rho (1 - D_r^i))D_p^i.
\end{equation}
In the simulation, growth rates were normalized to keep the total population in the system at $S N$.

In each generation, family $i$ contains offspring families whose numbers follow a Poisson distribution with a mean proportional to $g_i$. Offspring families inherit the parental family's $t$ and $p$ values with a slight mutation, which is represented by Gaussian noise with variance $\mu^2$. Then, all parental families are removed. Initially, $(t, p) = (0, 0)$ for all the families. The parameters involved in this model are the number of societies $S$, initial number of families in society $N$, mutation rate $\mu$, and relative importance of avoiding rivalry to cooperation $\rho$.

\diff{Note that, in reality, in-law cooperation is observed between kin groups united by actual marriage ties, and sexual rivalry occurs when two individuals actually prefer the same person. We approximated the former by the closeness of $t$ and $p$, since the mate's $t$ is close to $p$. Similarly, we approximated the latter by the overlap of $p$, i.e., the hypothetical mate preference. Suppose that people actually prefer several mate candidates before marriage based on $p$. As the attractiveness of people is unequal and people can prefer multiple mate candidates, the conflict over the same particular mates occurs. If two individuals prefer the same unmarried candidate, it will be mating competition, but if one prefers a married individual, it will be infidelity. As we mentioned above, sexual rivalry exists even when the sex ratio is equal \cite{schacht2014too, moss2016biased}, and be averted by marital rules \cite{chagnon2017manipulating}.}

There are other possible mathematical formulations for expressing kinship interactions. Here, we adopted the abovementioned ones to keep the growth rate positive and finite. However, the following results were qualitatively robust against the choice of mathematical formulations as long as kin and in-law cooperation and sexual rivalry were appropriately considered. For example, the simulation results are qualitatively unchanged when using other functions that decrease with distance, such as $\exp(-|x|)$ or $1/(1 + x)$, instead of the above formulation by using $\exp(-x^2)$. Similarly, by defining the growth rate as $b - d_c (1 - D_f^i) - d_m D_r^i$ or $b/((1 + d_c(1-D_f^i))(1 + d_mD_r^i))$, where $d_c$ ($d_m$) denotes the need for cooperation (avoidance of rivalry) and $d_m / d_c = \rho$, we obtain essentially identical results. Moreover, the evolved structures are almost identical by coupling families based on the marriage possibility defined by $\exp(-(t_i - p_j)^2)$, instead of calculating the degree of preferredness of each family as in the current model (see \cite{itao2020evolution, itao2022emergence}). They are also unchanged when we use the degree of preferredness $D_p' = \ev{\exp(-(t_i - p_j)^2) / \ev{\exp(-(t_k - p_j)^2)}_k}_j$ to measure the relative favorability of each woman in the society to a given man. In addition, this $D_p'$ is approximately equal to the current $D_p$ when kinship structures shown later evolve (see Supplementary Text).

Although we adopted the above formulation to make the model simple, some variants of the model considering the variations of kinship interactions produce essentially identical results. First, the in-law cooperation can be bidirectional, a brideprice from the groom's family to the bride's, or a dowry in the opposite direction \citep{leach1982social, anderson2007economics, shenk2011rebirth}. We modeled the brideprice-type interaction as it is the most frequent in the societies in the SCCS (though the dowry-type is the most frequent in terms of the population performing it) \cite{murdock1969standard, kirby2016d, anderson2007economics}. Alternatively, Figs. S1 and S2 show the simulation results where the dowry-type and bidirectional cooperation are modeled. They are essentially identical to the results to be presented below. Second, in the above model, sexual rivalry occurs when female preferences overlap. In reality, sexual rivalry is more common in men \cite{daly1982male, chagnon1988life}. Then, Fig. S3 shows the simulation results of a model where men prefer women and overlapping male preferences lead to sexual rivalry. The consistency between the following results and Fig. S3 suggests the generality of the findings regarding the evolution of kinship structures when sexual rivalry caused by overlapping mate preferences is modeled.


\section*{Results}
\subsection*{Numerical results}
\begin{figure}[t]
 \centering
  \includegraphics[width=\linewidth]{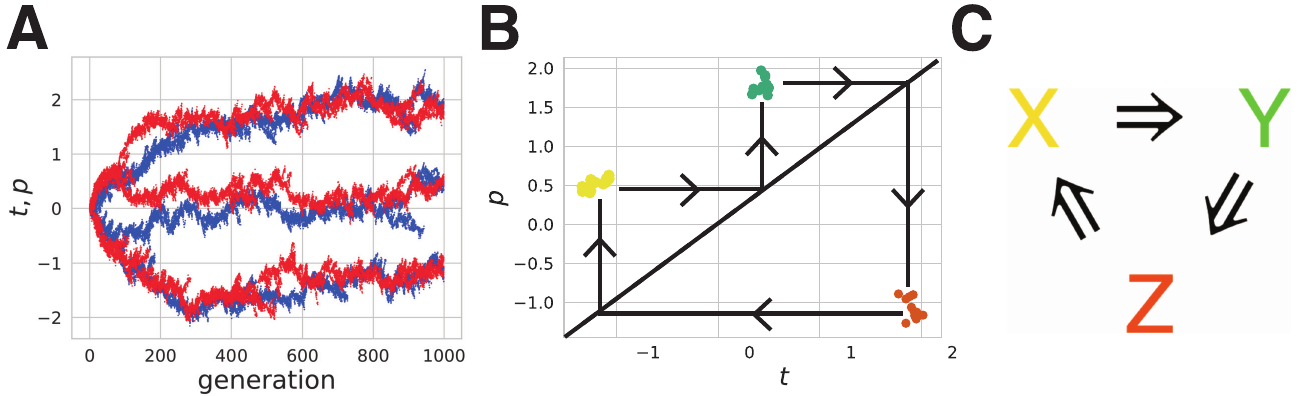}
\caption{
Example of evolved structures. (A) Temporal change of $t$ (blue) and $p$ (red) values within a society. They bifurcate into several branches with matching traits to someone's preferences. (B) Distribution of $t$ and $p$ in the final state. Clans X (yellow), Y (green), and Z (orange) prefer each other cyclically. The dashed line shows a $t = p$ line. (C) A schematic of the emergent structure.
}
\label{fig:clan_phys_generalized}
\end{figure}

Simulations were performed for $1000$ generations, within which stationary structures were obtained. (Note that the structures were formed in much fewer generations.) Fig. \ref{fig:clan_phys_generalized} illustrates an example of temporal change. Families' growth rates increase if they have more cooperators and mate candidates and fewer rivals. In other words, there is evolutionary pressure to gather $t$ and separate $p$ under the constraint that $t$ should match someone's $p$. Consequently, $t$ and $p$ values diverged to form several clusters to prevent sexual rivalry. Families in the same cluster are cultural siblings because of similar $t$ values and they have in-laws as additional cooperators because they prefer families in different clusters.
These emergent clusters are exogamous kin groups that can be considered clans in anthropology. Evolved clans require other clans to reproduce and their interdependent relationships constitute kinship structures. 

Families’ interactions depend on the Gaussian function of the distance between $t$ and $p$. Hence, when the distance between clans is $2$, the interaction is $e^{-4}\simeq 2\%$ of that with a distance $0$. In Fig. \ref{fig:clan_phys_generalized}, men in clan $X$ almost always marry women in clan $Z$, receive cooperation from members of clan $X$ and $Y$, and engage in rivalry with members of clan $X$. Thus, a preferential relationship can be written as $X \Rightarrow Y \Rightarrow Z \Rightarrow X$ with marriage cycle length $c_m = 3$. Here, we apply the $X$-means clustering method to optimize the number of clusters by adopting the Bayesian information criterion \cite{pelleg2000x}.

\begin{figure}[t]
 \centering
  \includegraphics[width=\linewidth]{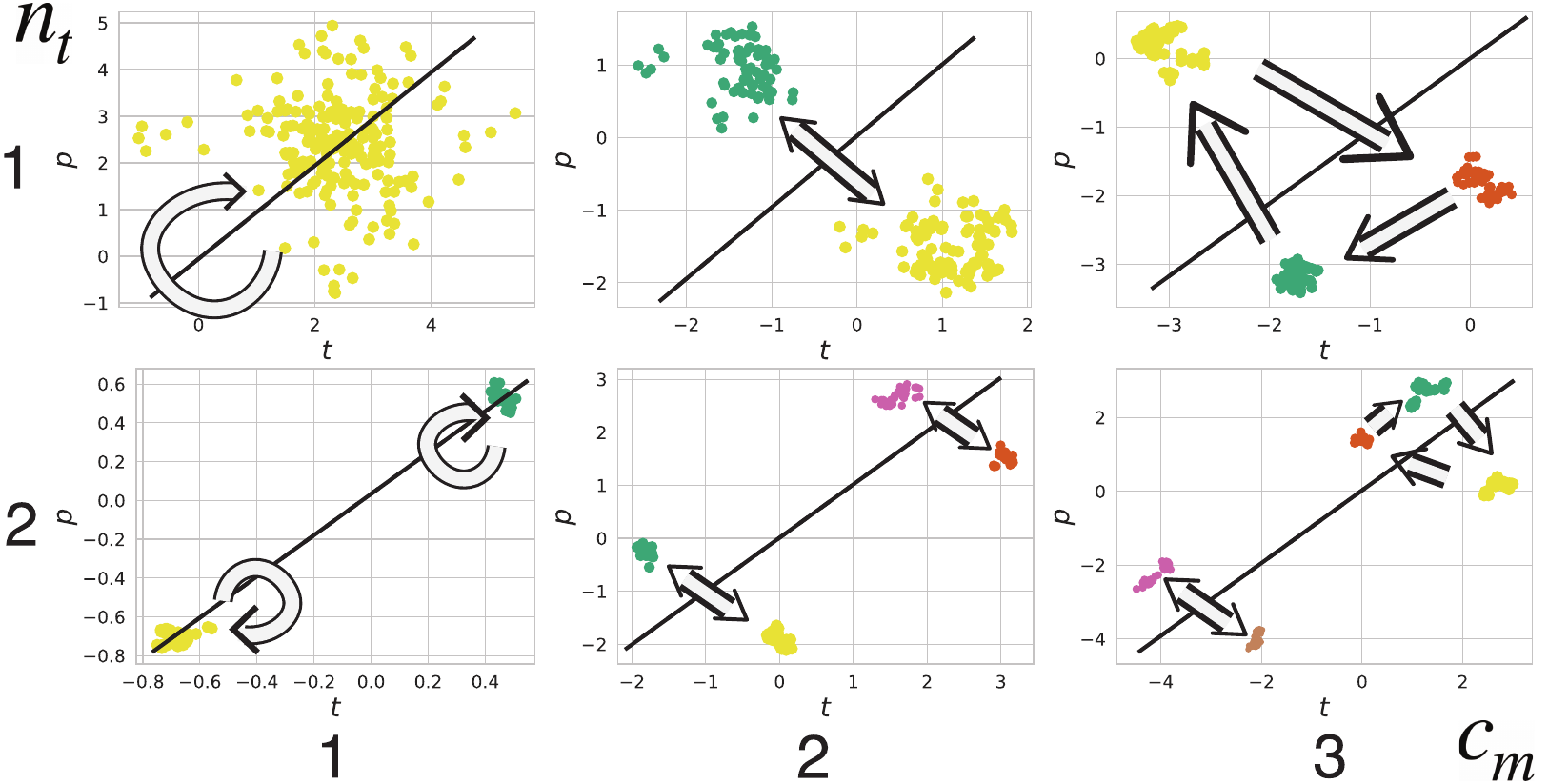}
\caption{
Examples of the evolved structures with different numbers of tribes $n_t$ and marriage cycle lengths $c_m$, plotted in the $t-p$ space. Parameters are set as $S = 100, N= 100$ and $\mu = 0.03$. $\rho = 0.1$ for the upper left and center with $(c_m, n_t) = (1, 1)$ and $(2, 1)$, and $\rho = 10$ otherwise.
}
\label{fig:clan_phys_examples}
\end{figure}

Diverse structures evolved in the simulation as shown in Fig. \ref{fig:clan_phys_examples}. To characterize them, we measured the marriage cycle length $c_m$, and the number of tribes $n_t$. The latter represents the number of distinct marriage cycles in a society. People move by marriage between clans in the same cycle and relatives belong to one clan of them. However, clans in different cycles have no demographic relationship, as if they are from different tribes.
The number of clans $n_c$ in society roughly equals $c_m n_t$, although it can differ if tribes have different cycle lengths (e.g., when a tribe with endogamy and one with dual organization coexist).
Hence, to have many clans to reduce sexual rivalry, either the marriage cycle or the number of tribes can increase. \diff{In Fig. \ref{fig:clan_phys_examples}, clans in the same society have similar population sizes. Thus, the degree of preferredness $D_p \simeq 1$, the density of cooperative friends $D_f \simeq 1 / n_c$ (if $c_m = 1$) and $\simeq 2 / n_c$ (otherwise), and the density of competing rivals $D_r \simeq 1 / n_c$. As $D_f$ and $D_r$ decrease and $D_p$ remains unchanged with $n_c$, structures with many clans evolve to avoid sexual rivalry by establishing strict rules to avert conflicts, which reduce cooperation.}

\begin{figure}[t]
 \centering
  \includegraphics[width=\linewidth]{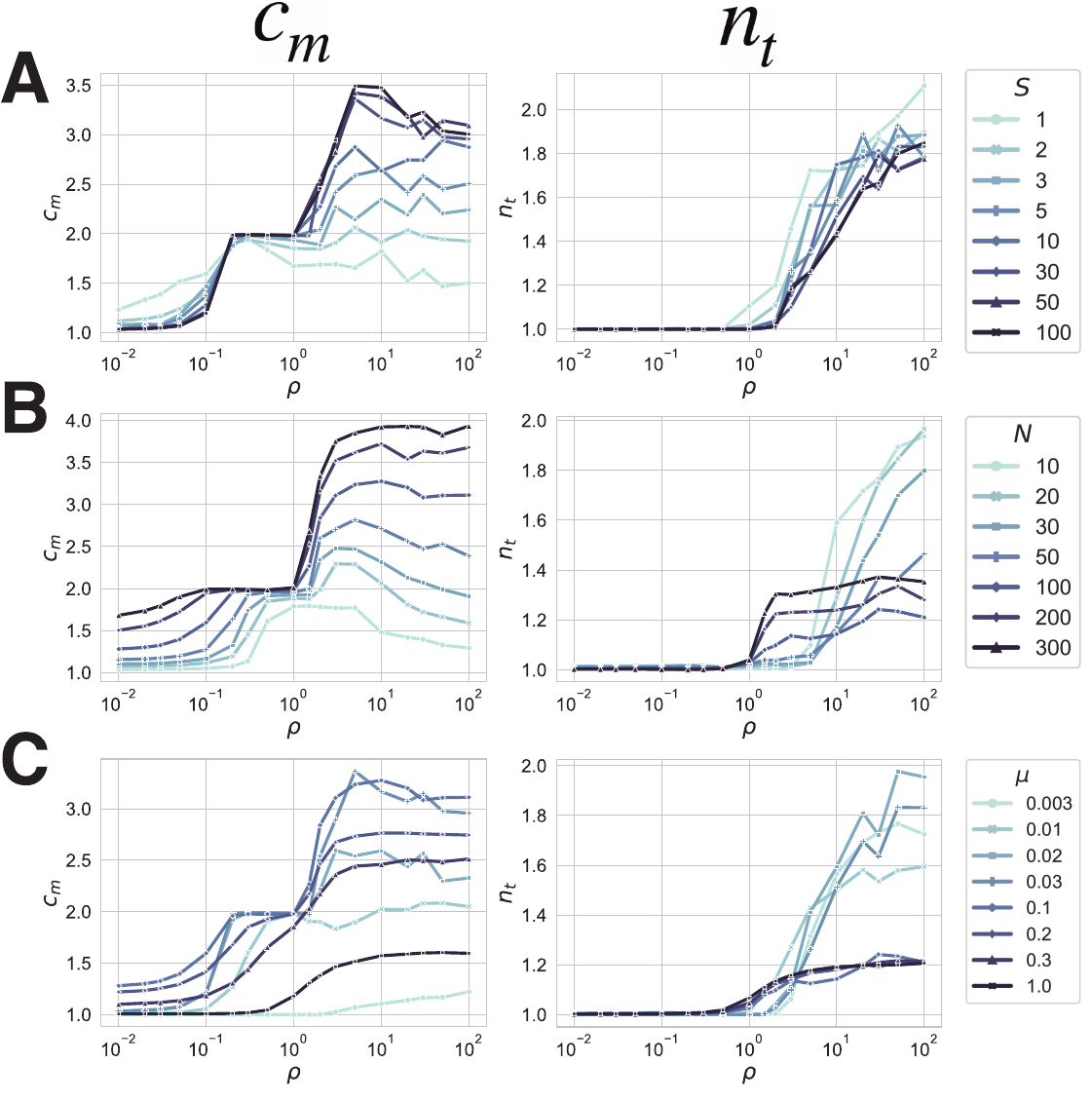}
\caption{
Dependence of the marriage cycle $c_m$, and the number of tribes $n_t$, on parameters representing (A) the number of societies $S$, (B) the number of families $N$, and (C) the mutation rate $\mu$. $S = 30,\ N = 100,$ and $\mu = 0.03$ unless explicitly stated. Figures show the average obtained from $100$ trials of simulation. Each simulation was performed for $1, 000$ steps.
}
\label{fig:clan_phys_phase_nsnfmu}
\end{figure}

The parameter dependencies of $c_m$ and $n_t$ are presented in Fig. \ref{fig:clan_phys_phase_nsnfmu}. As $\rho$ increases, $c_m$ and $n_t$ also increase. However, there are two plateaus at approximately $0.1 \le \rho \le 1$ and $\rho \ge 10$. The levels and ranges of the plateaus depend on $S,\ N$, and $\mu$. When the number of societies, $S$, is large, society-level selection operates strongly, and $c_m$ is large. It saturates when $S$ is sufficiently large. Hence, we set $S$ to a sufficiently large value, $100$, for the following analysis. Similarly, when the number of families in society, $N$, is large, $c_m$ is also large. However, the results for $\rho \ge 2$ depend significantly on $N$. As $\rho$ increases, $c_m$ increases for large $N$, whereas $n_t$ increases for small $N$. Finally, $c_m$ is the largest at the intermediate mutation rate $\mu$ $(= 0.03, 0.1)$ and $n_t$ increases only when $\mu$ is small. This suggests that, to have more clans, $n_t$ increases when $N$ and $\mu$ are small, whereas $c_m$ increases otherwise.

\begin{figure*}[bt]
 \centering
  \includegraphics[width=\linewidth]{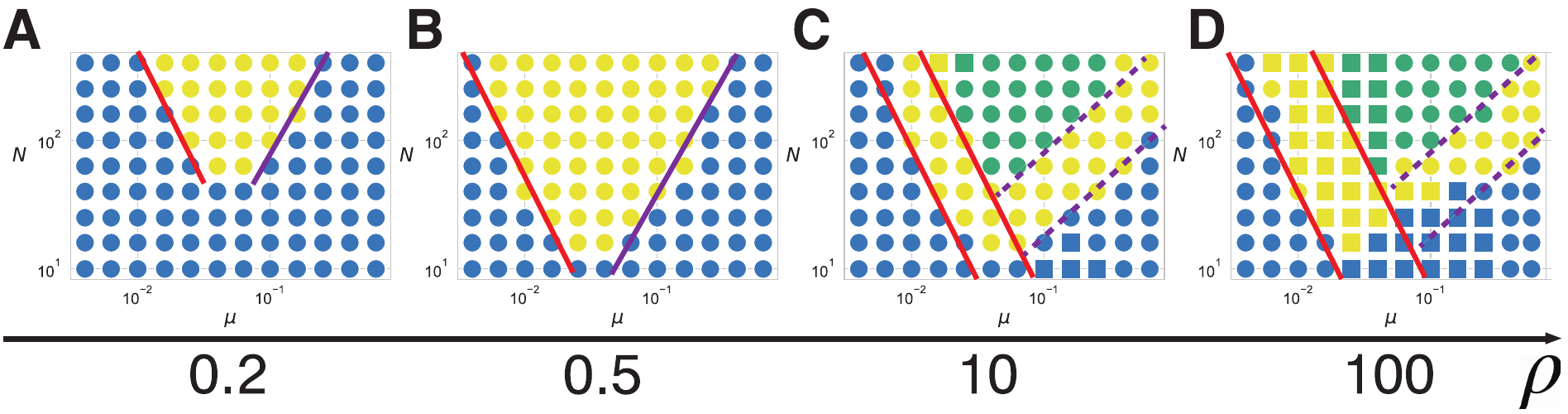}
\caption{
Phase diagram of the structures that evolve most frequently in $100$ trials of simulation under the given parameters representing the number of families, $N$, and the mutation rate, $\mu$, for different $\rho$. Colors indicate the marriage cycle length; endogamy, $c_m = 1$ (blue), dual organization, $c_m = 2$ (yellow), and generalized exchange, $c_m \ge 3$ (green). Shapes show the number of tribes; a single tribal society, $n_t = 1$ (circle), and multiple tribal society, $n_t \ge 2$ (square). Red lines show the isoclines of $N\mu^2$. Purple lines in (A, B) show isoclines of $\mu^2 / N$ and the dashed lines in (C, D) show those of $\mu / N$.
}
\label{fig:clan_phys_phase_r}
\end{figure*}

Fig. \ref{fig:clan_phys_phase_r} depicts the phase diagram of evolved kinship structures against $\mu$ and $N$ for different $\rho$. 
The phase diagrams are qualitatively distinguished into two types depending on $\rho$. For small $\rho$ ($= 0.2, 0.5$), the single-tribe dual organization (yellow circle) evolves if $\mu$ is within a certain range bounded by $N\mu^2$ and $\mu^2 / N$. This range expands for large $N$ and $\rho$. If $\mu$ is outside the range, single-tribe endogamy evolves (blue circle). By contrast, for large $\rho$ ($= 10, 100$), various structures evolve the boundaries of which are scaled by $N\mu^2$ and $\mu / N$.
These scaling laws in phase boundaries are further explored in a subsequent analysis.

\subsection*{Analytical estimation}
Here, we analytically estimate the dependence of $c_m$ and $n_t$ on $\rho,\  N$, and $\mu$. 
As we consider multilevel evolution, we first examine which kinship structures are adaptive at the society level. We then investigate the family-level dynamics that give rise to these structures in each society.

We first explain the different trends shown in Fig. \ref{fig:clan_phys_phase_nsnfmu} depending on whether $\rho$ is larger or smaller than $2$ by considering the society-level adaptiveness of kinship structures.
First, we calculate the growth rate in the ideal case by assuming that all families in the same clan have the same $t$ and $p$ values and that their values of mate clans match exactly. In this case, the growth rate of all families in society is the same, so it can be regarded as the population growth rate at the society level.
For instance, if $c_m = 2$, half of the families have $(t, p) = (-a, a)$ and the others have $(a, -a)$ with a constant $a > 1$. The cooperators are their own and their mates' clan members, and the rivals are their clan members. If $c_m = 1$, kin and mates overlap, whereas there are two cooperative clans for $c_m \ge 2$. Thus, the number of cooperative clans is given by $\min(2, c_m)$. Then, the growth rate of society $g = (D_f + \rho(1 - D_r))D_p$ is calculated as a function of the marriage cycle length $c_m$ and the number of clans in society $n_c$ ($\simeq c_m n_t$) as
\begin{align}
g(c_m, n_c) &= \min(2, c_m)/n_c + \rho(1 - 1 / n_c)  \\
&= \rho + (\min(2, c_m) - \rho)/n_c.    
\end{align}
Thus, for $c_m \ge 2$, $g(c_m, n_c) = \rho + (2 - \rho) / n_c$. Then, the optimal structure depends on $\rho$; if $\rho < 2$, the second term is positive, and society with smaller $n_c$ has a higher growth rate. Hence, the optimal structure is $c_m = 2$ and $n_t = 1$ (note that $n_c$ cannot be $1$ because $c_m \ge 2$). In contrast, if $\rho \ge 2$, the second term is negative, and larger $n_c$ is adaptive. 
For structures with $c_m = 1$, there is at least one structure with $c_m \ge 2$ that exhibits higher fitness. If $c_m = 1$ and $n_c \ge 2$,  $g(1, n_c) = \rho + (1 - \rho) / n_c < \rho + (2 - \rho) / n_c =  g(c_m \ge 2, n_c)$. If $c_m = n_c = 1$, $g(1, 1) = 1 < 1 + \rho /2 =  g(2, 2)$. Therefore, the optimal structure is $c_m = 2$ and $n_t = 1$ if $\rho < 2$, and otherwise, $n_c (\simeq c_m n_t) = \infty$.

This dependence on $\rho$ can be explained by considering the effective force operating between families in the $t-p$ space.
For $\rho \ge 2$, the growth rate increases when a family can reduce one rival, even if it simultaneously loses one kin and in-law. Hence, it is adaptive for families to distance themselves from other families in the $t-p$ space. (Note that families must be preferred to find mates; $t$ and $p$ should match between a pair of families. Hence, it is impossible to have a small variation of $t$ and a large variation of $p$ simultaneously.) This indicates a repulsion between families. By contrast, when $\rho < 2$, the attraction operates to gather families in the $t-p$ space.

Thus far, we have discussed the optimal structure by ignoring fluctuations.
It is necessary to examine whether such a structure is explorable and maintainable by considering the intra-societal dynamics that generate each structure.
In contrast to the above analysis, various structures (including less adaptive ones) evolved in the simulation. To explain the phase diagram in Fig. \ref{fig:clan_phys_phase_r}, we need to consider the family-level dynamics driven by the fluctuation due to $N$ and $\mu$.

First, we investigate the transient dynamics from clan endogamy ($c_m = 1$) to dual organization ($c_m = 2$). Let us assume that a fraction $q$ of families form dual organization (clans A and B) and $1-q$ performs endogamy (clan C). The cooperators of families in clan A are those in A and B, and their rivals are in A. Similarly, the cooperators and rivals of families in C are those in C. Then, families in clans A and B have a growth rate of $g_d$ and those in C have $g_e$.
Each growth rate is given by
\begin{align}
    g_e &= (1-q) + q \rho\\
    g_d &= q + (1 - q /2) \rho.
\end{align}
Subsequently, the temporal change in $q$ is given by $\dot{q}(q) = g_d - g_e = (2 - 3 \rho / 2)q + \rho - 1$.
If $\rho < 1$, $\dot{q} < 0$ for $q < q^\ast = \frac{1}{2} - \frac{\rho}{8-6 \rho}$ and $\dot{q} > 0$ otherwise.
Thus, when $\rho < 1$, clan endogamy is a locally stable attractor, and dual organization emerges if more than a fraction $q^\ast = \frac{1}{2} - \frac{\rho}{8-6 \rho}$ of families happen to form mate clans with a matched $t$ and $p$ by fluctuations. Therefore, we need to consider the probability of such a formation depending on the fluctuations in the system. We call this the overall fluctuation, whose magnitude is determined by $N$ and $\mu$.
By contrast, $\dot{q}(q)$ is always positive when $\rho \ge 1$. Thus, clan endogamy is unstable and multiple clans are organized spontaneously. Complex adaptive structures’ evolution depends on the probabilities of their formation and maintenance, which are determined by $N$ and $\mu$ as discussed later. (Note that the optimal structures change at $\rho = 2$ while the local stability of the clan endogamy changes at $\rho = 1$.)

Moreover, under large fluctuations, the traits and preferences of mate clans may not match appropriately, leading to a decline in cooperation and marriage possibilities, and consequently to the collapse of interdependent relationships.
For example, in dual organization, let us assume that half of the families have $(t, p) = (-a, a)$ and the others have $(a + \delta, - a - \delta)$ with constants $a > 1$ and $\delta > 0$. Then, the degree of preferredness of each family decreases to $D_p^i = \exp(-\delta^2)$ and its density of friends to $D_f^i = 1 + \exp(-\delta^2)$, whereas its density of rival is unchanged. These declines lead to smaller fitness.
We term a fluctuation causing such a mismatch of mate clans as marital fluctuation. Its dependence on $N$ and $\mu$ is further explored in the following analysis.

\begin{figure}[tb]
 \centering  \includegraphics[width=\linewidth]{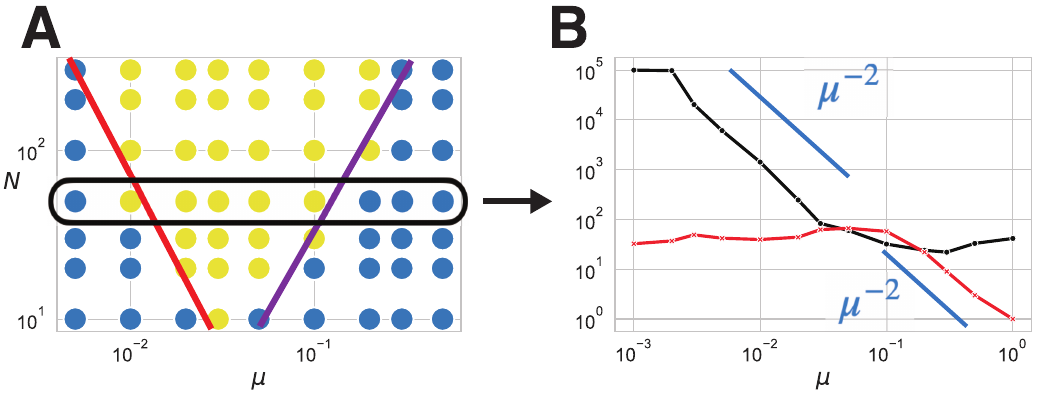}
\caption{
Explorability and maintainability of dual organization depending on the mutation rate $\mu$ for the case where the attraction is dominant ($\rho = 0.5$). (A) Phase diagram of kinship structures (the same as Fig. \ref{fig:clan_phys_phase_r}(B)). (B) Time for formation and collapse of dual organization when $S=1$.
The black line shows the average generation required to form dual organization from the initial homogeneous state. The red line shows the average generation required for the collapse of the dual organization into endogamy.
}
\label{fig:clan_phys_formation_maintenance}
\end{figure}

Next, we examine the phase diagram in Fig. \ref{fig:clan_phys_phase_r}(A, B) for $\rho < 1$. As shown in the figure, the boundaries are defined as the isoclines of $N\mu^2$ and $\mu^2 / N$. The possible structures for $\rho < 1$ are dual organization ($c_m = 2$) and clan endogamy ($c_m = 1$). As noted above, both are locally stable, whereas dual organization is $1 + \rho / 2$ times more advantageous than endogamy at the society level. We then explore typical times for the formation and collapse of dual organizations.
Fig. \ref{fig:clan_phys_formation_maintenance} indicates that, as $\mu$ increases, the generation required for formation of dual organization decreases first and that for collapse decreases second. (Here, $S$ is set to $1$ to focus on the intra-societal dynamics. If $S > 1$, the formation time will be shortened, because once a society reaches dual organization, it will spread to other societies because of society-level selection.)
Thus, for a small $\mu$, dual organization is difficult to generate, whereas for a much larger $\mu$, it is difficult to maintain. Dual organization evolves only for a range of $\mu$ the boundaries of which are scaled by $N\mu^2$ and $\mu^2/ N$.

First, the left boundary is explained by considering the transient dynamics required to form dual organizations.
Because $t$ and $p$ values are inherited with a slight mutation with variance $\mu^2$, the fluctuations added to the entire t-p distribution of $N$ families have a magnitude scaled by $N \mu^2$. Thus, dual organization emerges when the overall fluctuation, whose magnitude is proportional to $N \mu^2$, is larger than a certain constant, $\alpha$, to obtain the fraction, $q \ge q^\ast$, of the families that form it.
Next, we consider its maintenance to derive the right boundary scaled by $\mu^2/ N$:
Given that dual organization is an interdependent structure, the growth rates of families vary due to the above marital fluctuations. As attraction operates among families in the $t-p$ space when $\rho < 1$, families gather by attraction and diverge because of mutations. Thus, the magnitude of the mismatch of mate clans depends on a random motion in the $t-p$ space of the centroids of families in the same clan. Now, each family fluctuates with a magnitude of $\mu^2$, and thus, the centroids fluctuate with a magnitude of $\mu^2 / N$.
This mismatch rate $\mu^2 / N$ should be smaller than a certain constant $\beta$ for dual organization to be maintained.
In summary, dual organization evolves when $\alpha / N < \mu^2 < \beta N$. As $\rho$ increases, dual organization becomes more advantageous than clan endogamy, and $q^\ast$ decreases, leading to smaller $\alpha$ and larger $\beta$, that is, the larger regions in the phase diagram. Fig. \ref{fig:clan_phys_phase_r} indicates that $(\alpha, \beta) \simeq (4 \times 10^{-2},\ 1 \times 10^{-4})$ and $(1 \times 10^{-2},\ 4 \times 10^{-4})$ for $\rho = 0.2$ and $0.5$, respectively.

\begin{figure}[tb]
 \centering
  \includegraphics[width=\linewidth]{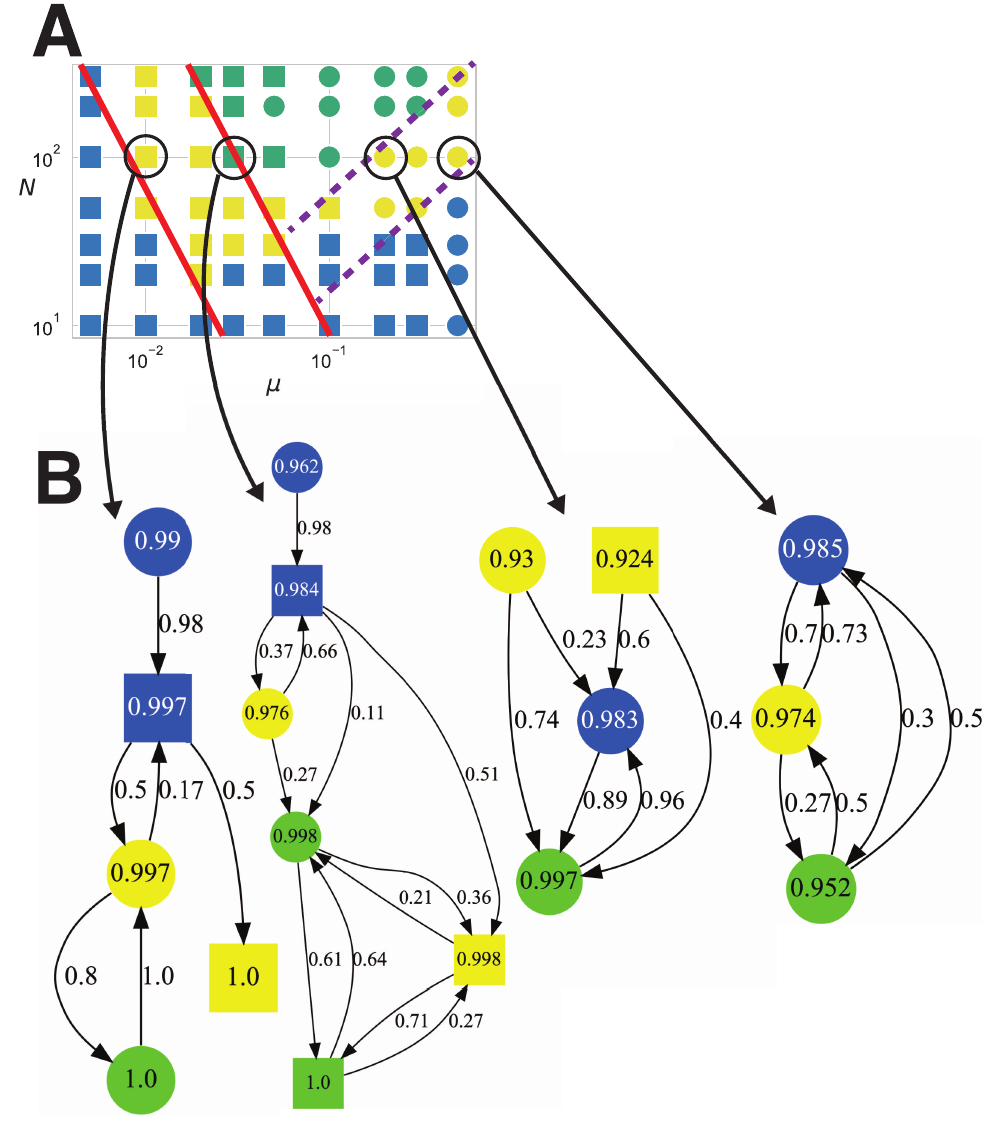}
\caption{
State transition probability of structures for the case where the repulsion is dominant ($\rho = 100$). (A) Phase diagram of kinship structures (the same as Fig. \ref{fig:clan_phys_phase_r}(D)). (B) The state transition probabilities of the system. Values in the nodes show the probability of staying and those on the edges show the relative probability of transition. Edges with weights greater than $0.1$ are shown.
}
\label{fig:clan_phys_transition}
\end{figure}

Next, we consider the case where $\rho \ge 2$. The boundaries in Fig. \ref{fig:clan_phys_phase_r}(C, D) are scaled by $N\mu^2$ and $\mu / N$. Now, structures with many clans are preferred. Thus, the possible structures are endogamy, dual organizations, and generalized exchanges ($c_m \ge 3$), in which societies can have multiple tribes.
A detailed analysis is more difficult than in the above case with $\rho < 1$ because there are many possible structures.
Still, when $N$ is finite, several families must happen to have matched $t$ and $p$ values to formulate complex structures with many clans. Therefore, the fluctuations drive the transition between locally stable structures. Fig. \ref{fig:clan_phys_transition} shows the state transition probabilities between structures. A high probability of remaining in each structure (especially under small $\mu$) indicates local stability. Similar to the case where $\rho < 1$, the probability of generating complex structures increases with the overall fluctuation, $N \mu^2$.
As $N \mu^2$ increases, structures with larger $c_m$ evolve and the boundary is almost independent of $\rho$.
For larger $\rho$, societies have many tribes to avoid sexual rivalry. 
However, when $\mu$ is too large, families move largely in $t-p$ space. This increases the probability that families from different tribes take similar $t$ and $p$ and collide with each other. This greatly reduces their growth rate due to strong rivalry or low marriage possibilities. Thus, a single-tribal society evolves for a large $\mu$.

Structures collapse owing to marital fluctuations, that is, mismatches between $t$ and $p$ as in the case with $\rho < 1$. However, the repulsion between families in the $t-p$ space causes families to be widely distributed within a cluster whose radius is proportional to $\mu$. The mismatch rate is then determined based on whether there are families that prefer each family, which is estimated by $\mu n_c / N$ (rather than
based on the distance between the cluster centers' $t$ and $p$). As $\mu / N$ increases, the number of clans decreases because of a larger mismatch.
The boundaries in Fig. \ref{fig:clan_phys_phase_r} are scaled by $N\mu^2$ and $\mu / N$, which is consistent with the current analysis.

\section*{Discussion}
In this study, we built a model of kinship interactions, including kin and in-law cooperation and sexual rivalry.
The simulation results demonstrate that these interactions lead to the formation of clans as cultural kin groups, with cyclic marriage preferences. This illustrates the evolution of kinship structures through the spontaneous formation of interdependent groups. Differentiation into clans results from the need to avoid rivalry.
Interdependency between clans arises from the need to have another cooperator as in-law rather than as kin.
The evolved structures are characterized by the marriage cycle length $c_m$ and number of tribes (as distinct cycles) $n_t$. As the relative importance of avoiding rivalry versus cooperation, $\rho$, increases, $c_m$ or $n_t$ increases. Evolved structures depend on the fluctuation in traits and preferences and mate access, which is determined by population size $N$ and mutation rate $\mu$. 

The branching pattern of traits and preferences in Fig. \ref{fig:clan_phys_generalized}(A) resembles the ``segmentary lineage system'' which is characterized by branching relationships among cultural groups, where each clan or lineage traces its origin back to a common ancestor. Groups form alliances or engage in conflicts based on their perceived distance from each other within the branching pattern. Such a system is observed, for example, in African and Arab societies \citep{evans1940nuer, sahlins2022segmentary}.

In this model, $\rho$ determines the attraction or repulsion between families in the $t-p$ space, as well as optimal structures and scaling laws in the phase diagram. When $\rho < 2$, attraction operates, and dual organization (i.e., two clans preferring each other) is optimal. In contrast, when $\rho \ge 2$, repulsion occurs, and structures with as many clans as possible are favored. In any case, structures shift between several locally stable attractors driven by fluctuations in the $t-p$ space, whose explorability is scaled by the overall fluctuation $N\mu^2$, determined by the number of families in society $N$, and mutation rate $\mu$. However, complex structures with many clans collapse because of marital fluctuations caused by a mismatch between $t$ and $p$ of mate clans. When attraction operates, the probability of collapse is given by $\mu^2 / N$, which represents the mismatch of the gravity centers of clans in the $t-p$ space. In contrast, when repulsion occurs, the probability of collapse is given by $\mu / N$, which represents the overlap of families distributed in the $\mu$-radius.
The number of societies, $S$, determines the strength of society-level selection. When $S$ is sufficiently large, its strength saturates.

These scaling laws suggest that throughout human history, as society has grown and population size has increased, more complex structures have emerged. Demography is considered a major determinant in the maintenance of cultural complexity \cite{powell2009late}. This study introduces another complexity class: the interdependency of differentiated groups. It is shown that not only must $N$ be large but also that $N$ in combination with $\mu$ determines the explorability and maintainability of the structures. Significantly, we considered the various model variants in which dowry-type or bidirectional in-law cooperation or mate preference by males are incorporated. Their results illustrated in Figs. S1–S3 consistently demonstrate the existence of distinct phases of kinship structures and associated scaling laws concerning $N$ and $\mu$. 
Thus, the current findings are considered general consequences of kin and in-law cooperation and sexual rivalry, regardless of their specific forms\footnote{In statistical physics, phenomena such as the existence of distinct phases (gas, liquid, and solid) and scaling laws for phase boundaries often exhibit universality across various models. In this respect, it will be reasonable to expect that the model variants will demonstrate consistent results when the essential characteristics are appropriately modeled. In the current case, as long as clans are organized and individuals cooperate within their own and their mates' clans, while engaging in rivalry within their own clans, the outcomes remain unaffected by either the direction or the gender of those involved in the interactions.}.
(Note that in reality, when the population size increases very much, kinship loses its significance as the organizing principle of societies because of the development of other interactions, such as gift-giving and other political, social, and/or religious ones \cite{itao2023transition, itao2023emergence}.)

We also conducted empirical tests of our theoretical results by analyzing data related to population size, parental influence in mate selection, the necessity for cooperation, and the need to avoid rivalry. These analyses were aimed at roughly estimating the parameters $N$, $\mu$, and $\rho$. Due to potential arbitrariness in variable selection and data availability limitations, the estimations presented in Fig. S4 should be interpreted as indicative of qualitative trends rather than precise parameter dependencies of kinship structures. The observed dependency of $c_m$ on these parameters aligned with our theoretical predictions, though it lacked the detailed elaboration necessary to empirically demonstrate the scaling laws.
Furthermore, it was not feasible to empirically examine the dependency of $n_t$. These challenges remain the limitations of our current study.

In the current model, we simplified $t$ and $p$ as scalar variables and considered unilineal inheritance only. In real society, these traits can be inherited both paternally and maternally \cite{Goody1961, levi1965future}. Indeed, in the SCCS (the variable 70), 40\% of societies are patrilineal, 14\% are matrilineal, 5\% are double descent (referring to both lines), 3\% are ambilineal (referring to either line in each generation), and 37\% do not comprise a corporate kin group \cite{murdock1969standard, kirby2016d}.
When both patrilineally and matrilineally inherited traits are significant in classifying children's attributions, children can have different (or mixed) attributions from their parents \cite{romney1958simplified}. In our previous study, we built an extended model with two-dimensional $\bm{t}$ and $\bm{p}$, where the first-dimensional components were inherited paternally and the second-dimensional components were inherited maternally, and demonstrated the evolution of a further variety of kinship structures \cite{itao2020evolution, itao2022emergence}. Nonetheless, essential mechanisms were investigated in this study.

Because our interest is in cultural kinship, we have ignored genetic relationships between families. However, the empirical study has reported that depending on the type of cooperative behavior, either genetic or cultural kinship is important \cite{alvard2009kinship}. Genetic relatedness varies with clan size, and thus, the development of cultural ideology to knit clans may also depend on it. Such an interaction between genetic and cultural kinship remains a limitation of this study.

Our model shares some similarities with the biological models of sympatric speciation in that avoidance of competition leads to group formation. The differentiation of endogamous groups (i.e., $c_m = 1$) results from evolved mating preferences \cite{dieckmann1999origin} or resource competition \cite{kaneko2000sympatric}.
Conversely, humans develop the ability to recognize kin, leading to the organization of an affinal network of groups through exogamy ($c_m \ge 2$) \cite{chapais2009primeval, planer2021towards}. Our model includes cooperation between in-laws and thus exhibits the emergence of diverse kinship structures with interdependency rather than the mere divergence of groups.

However, the current mechanism could apply to broader interdependent structures where reproductive assistance is reciprocated by an increase in fitness (e.g., by feeding). In many complex biological systems, the origins of such structures are studied using multilevel evolution theory \cite{takeuchi2017origin, zadorin2023multilevel, nowak2010evolution, hogeweg1994multilevel}. Scaling laws have been extensively studied in multilevel evolutionary systems. Previous studies revealed that altruistic behavior ceases when $N\mu^\alpha$ is sufficiently large, where $\alpha$ is $1$ or less \cite{kimura2006diffusion, takeuchi2022scaling}.
Recall that in our model, the interdependent structure evolves when $N\mu^2$ is sufficiently large and collapses when $\mu^2 / N$ (or $\mu/N$) is large.
Our model considered the interactions between elements and the resultant interdependent structures, instead of mere altruism. This difference led to different scaling laws.
Subsequently, the scaling laws and their dependence on complexity class (altruism or interdependency) and operating force (attraction or repulsion) will be measured experimentally.

In conclusion, by proposing a simple model of kinship interactions, we demonstrate the spontaneous formation of cultural associations with cyclic mating preferences. As avoiding rivalry is relatively more important than cooperation, the marriage cycle length and number of tribes increase. The explorability and maintainability of complex structures depend on the overall fluctuations given by $N \mu^2$ and the marital fluctuations scaled by $\mu^2 / N$ (or $\mu / N$), respectively.
This study proposes the first possible causal mechanism of typical human social structures by referring to ethnographic observations, which should be empirically validated and compared with other potential explanations to enhance credibility as the causal explanation in future studies. 
It expands the scope of cultural evolution, sociophysics, and complex systems theory and offers a theoretical framework for the social sciences from a statistical physics perspective.

\matmethods{
\subsection*{Algorithms of the model}
The algorithm for population changes in families is as follows: For the family $i$ and its offspring family $i^\ast$, 
\begin{align}
D_f^i=&\langle\exp(-(t^i - t^j) ^2) \label{clan_phys_eq:friend} \\
&\ \ \ + (1 - \exp(-(t^i - t^j) ^2))\exp(-(p^i - t^j) ^2)\rangle_j,  \nonumber \\
D_r^i=&\ev{\exp(-(p^i - p^j) ^2)}_j, \label{clan_phys_eq:rival}\\
D_p^i=&\ev{\exp(-(t^i - p^j) ^2)}_j / \ev{\exp(-(t^i - t^j) ^2)}_j, \label{clan_phys_eq:preferred}\\
g^i =& (D_f^i + \rho (1 - D_r^i))D_p^i, \label{clan_phys_eq:fitness} \\
\intertext{For the offspring families of $i$, their $t$ and $p$ values are}
t^{i^\ast} =& t^i + \eta,\  p^{i^\ast} = p^i + \eta. \label{clan_phys_eq:inheritance}
\end{align}
Here,$\eta$ is independently sampled for each offspring family $i^\ast$. $\ev{\cdot}_j$ represents the expected value averaged over families in the same society with varying $j$. The number of the offspring families of $i$ follows Poisson($g^i / g$), where $g$ denotes the average of $g^i$ over all families in all societies.

The population growth of each family depends on the density of friends $D_f$, rivals $D_r$, and the degree of preferredness $D_p$, each given by Eqs. (\ref{clan_phys_eq:friend}-\ref{clan_phys_eq:preferred}), respectively. The number of offspring families of $i$ follows the Poisson distribution whose average is proportional to $g^i$ is given by \eqref{clan_phys_eq:fitness}. Here, the average is normalized as $g^i / g$ to fix the total population size in the system to $SN$.
Finally, each offspring family inherits parental families' $t$ and $p$ values with slight mutation, as shown in \eqref{clan_phys_eq:inheritance}. Here,$\eta$ is independently sampled for each offspring family. Then, all parental families die and the generation is altered.

}

\subsection*{Data Availability}
Source codes for the model can be found here: \url{https://github.com/KenjiItao/clan.git}.
\showmatmethods{}

\acknow{
The authors thank Koji Hukushima, Tetsuhiro S. Hatakeyama, Heidi Colleran, and Kim Sneppen for the stimulating discussions.
This research was supported by a Grant-in-Aid for Scientific Research (A) (20H00123) from the Ministry of Education, Culture, Sports, Science, and Technology (MEXT) of Japan; JSPS KAKENHI Grant Number JP21J21565 (KI); and Novo Nordisk Fonden Grant Number NNF21OC0065542 (KK).
}

\showacknow{}

\end{document}